\documentclass[fleqn,12pt,twoside]{article}
\usepackage{espcrc1}
\usepackage{graphicx}
\usepackage{epsfig}
\usepackage{subfigure}

\title{Neutron Captures in the r-Process -- Do We Know Them and Does It
Make Any Difference?}

\author{T. Rauscher\address{Department of Physics and Astronomy,
University of Basel,
4056 Basel, Switzerland}\thanks{Supported by the Swiss NSF 
(grants 2024-067428.01, \hbox{2000-061031.02}).}}

\begin{document}
\maketitle


\section{INTRODUCTION}
Nucleosynthesis of elements beyond the iron peak requires reactions with
neutrons due to the high Coulomb barriers which prevent charged particle
reactions. Approximately
half of the intermediate and heavy elements are created in the
r--process with neutron number densities $n_\mathrm{n}>10^{22}$
cm$^{-3}$, effective neutron energies around 100 keV, and short
process times of up to a few seconds. These conditions point to an explosive
site but the actual site has yet to be identified.
Self-consistent SNII models show persistent problems in explaining 
r--nucleosynthesis. In consequence, most r--process investigations focus on
simplified, parameterized models which allow to study the required
conditions and their sensitivities to nuclear inputs.
Due to the high neutron densities the r--process synthesizes very
neutron-rich nuclei far off stability which subsequently decay to
stability when the process ceases due to lack of neutrons or low
temperatures. This raises
the question whether we can predict reactions far off stability
sufficiently well to make statements about r--process conditions.

\section{UNCERTAINTIES IN REACTION RATES FAR FROM STABILITY}
There are two main problems in predicting nuclear cross sections far
from stability. The first concerns the prediction of the nuclear
properties needed as inputs to the reaction models, i.e.\ of nuclear
structure far from stability with all its uncertainties \cite{desrau04}.
The second problem is to identify the relevant reaction mechanism and possible
interplay between different mechanisms. With decreasing neutron
separation energy $S_\mathrm{n}$, direct neutron capture becomes more and more
important relative to compound capture and below a certain level density
the Hauser-Feshbach statistical model for compound capture
cannot be applied anymore \cite{rtk,nptom}.
However, the situation is
not grave since it is not necessary to know the rates directly in the
r--process path. It is a misconception to view
the formation of r--isotopes as a sequence
of neutron captures and $\beta$--decays, similar to an s--process
but proceeding further out from stability. In fact, at high temperature
$T$ and high $n_\mathrm{n}$ all neutron captures and
photodisintegrations occur faster by several orders of magnitude than
any $\beta$-decay in a given isotopic chain \cite{nptom}.
Since forward and backward rates are related by detailed balance, 
the cross sections cancel
out and the ratio is mainly depending on $S_\mathrm{n}$, $T$, and
$\rho$. Neutron captures will only start to matter during freeze-out
when the lifetimes become longer due to lower $T$ and lower
$n_\mathrm{n}$. It has been shown that the
freeze-out proceeds very quickly for realistic conditions \cite{frei99}.
On one hand
this limits the importance of neutron captures, on the other hand it
validates the investigations which were performed using approximations
such as instantaneous freeze-out \cite{kl}.
\begin{figure}[t]
\begin{center}
\mbox{
\subfigure{
\includegraphics[width=4.5cm,angle=-90]{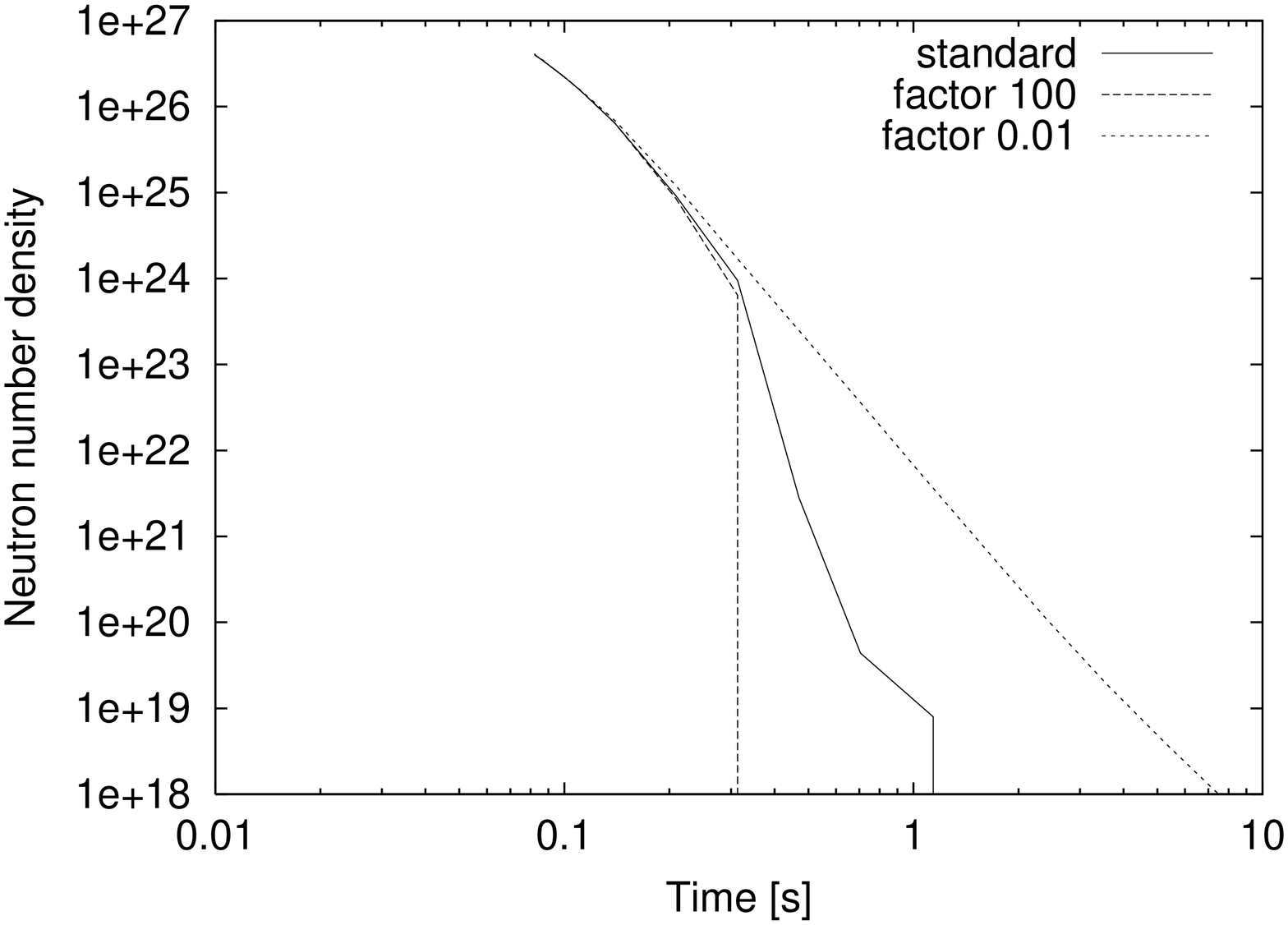}
}
\subfigure{
\includegraphics[width=4.5cm,angle=-90]{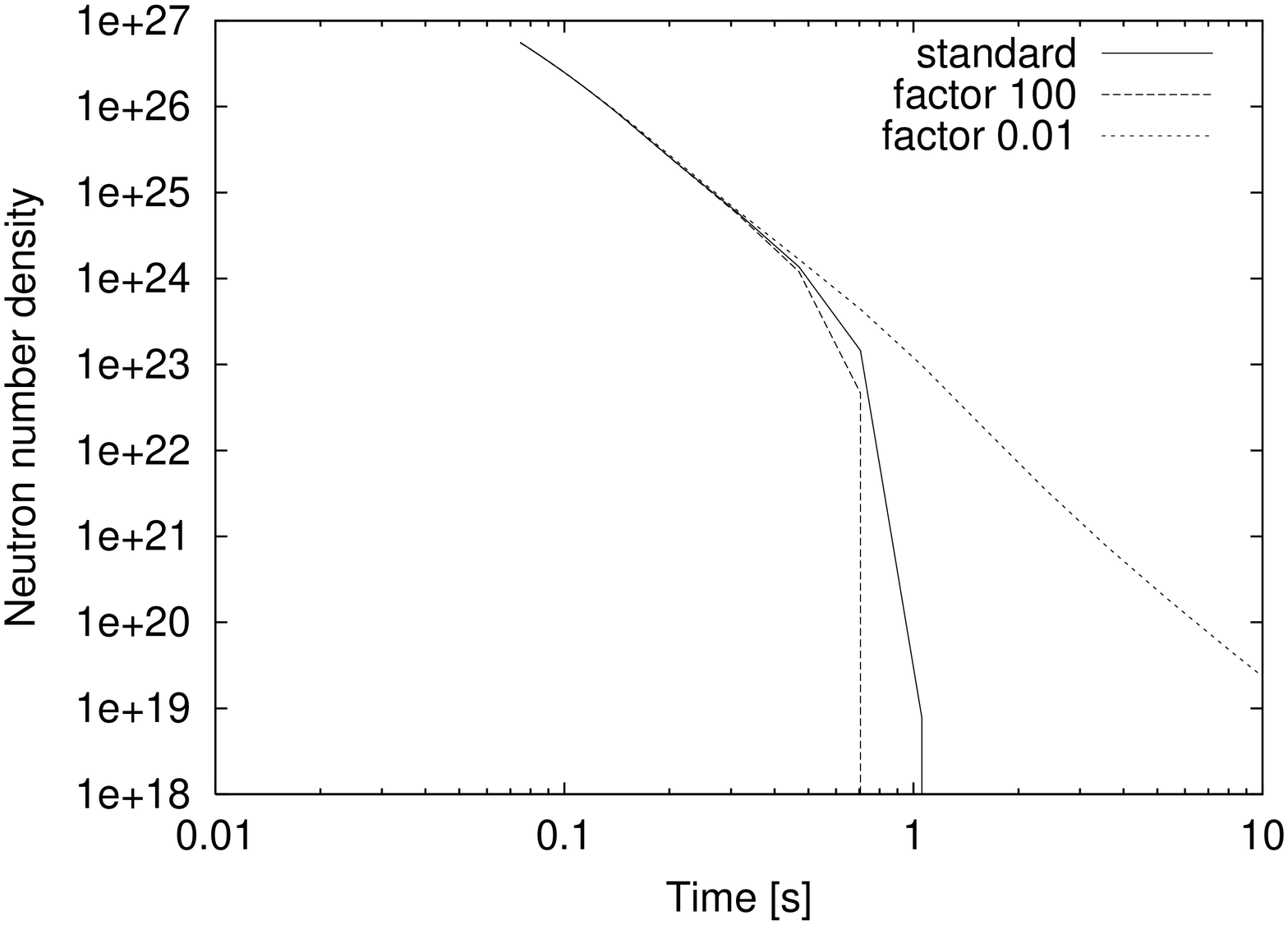}
}
}
\mbox{
\subfigure{
\includegraphics[width=5cm,angle=-90]{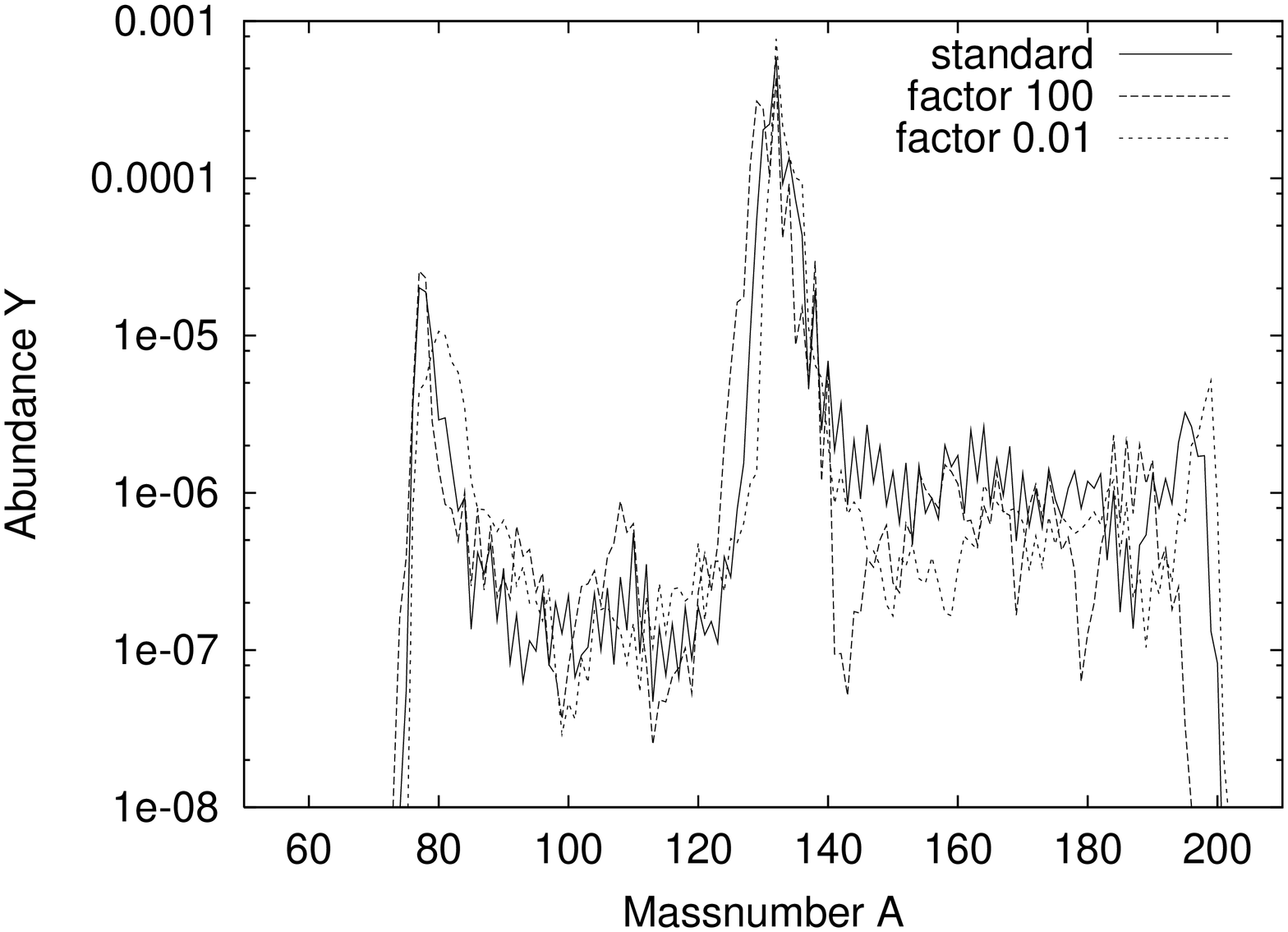}
}
\subfigure{
\includegraphics[width=5cm,angle=-90]{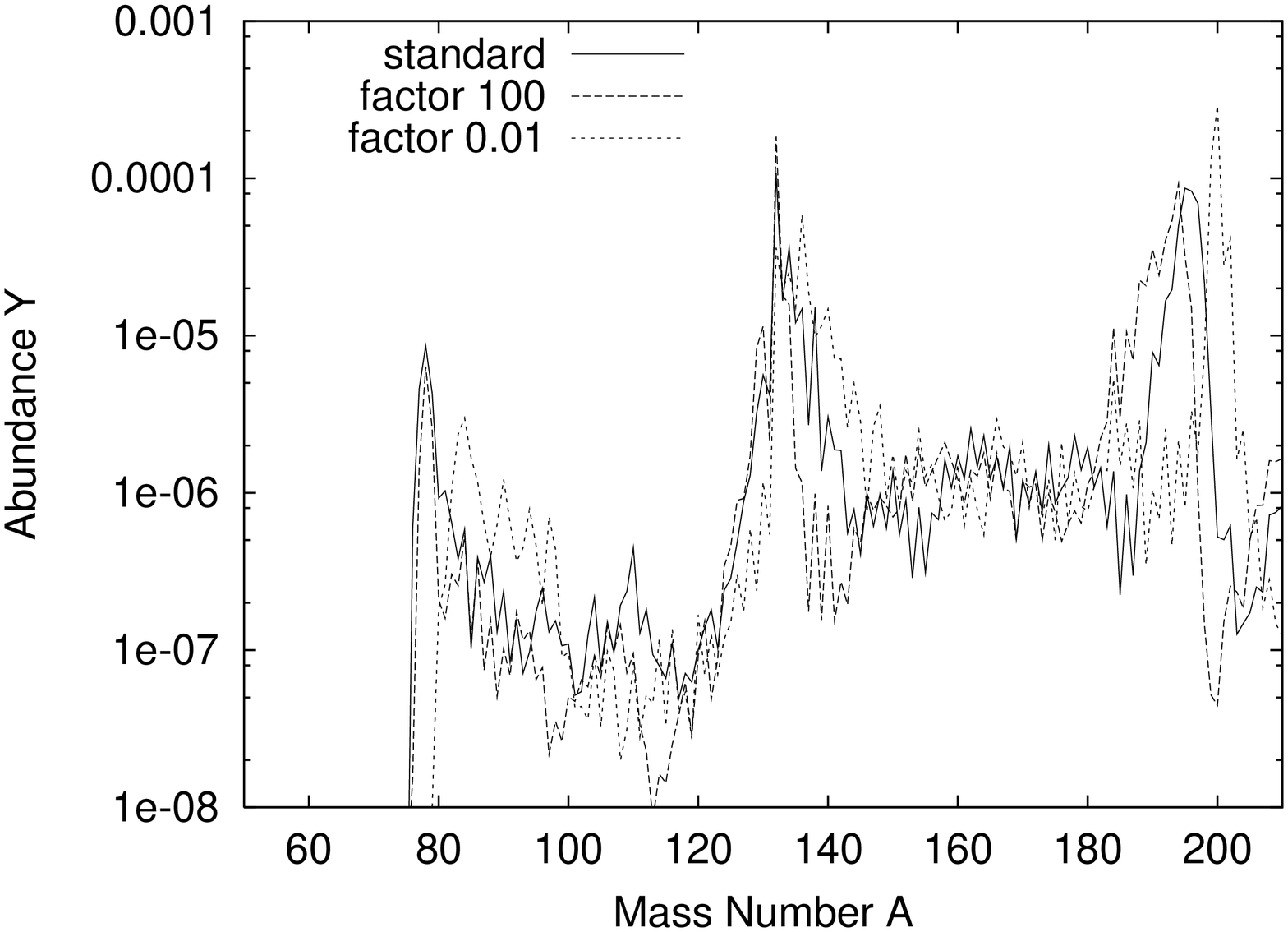}
}
}
\caption{\footnotesize Time evolution of neutron number densities (top,
starting at $t_\alpha$) and resulting
abundances (bottom) for $S=240$ (left) and $S=310$ (right). The neutron
rates were multiplied by factors 1 (full line), 100 (dashed), and 0.01
(dotted), respectively.}
\label{fig:nnabun}
\end{center}
\end{figure}

\section{ADIABATICALLY EXPANDING HOT BUBBLES}

Studying neutron captures in the freeze-out necessitates
dynamic r--process simulations. In this work
calculations in the model of an adiabatically expanding hot bubble were
performed similar to \cite{frei99}, combining a charged-particle
and an r--process network, but with updated, temperature-dependent rates
\cite{rath00}, thus improving a previous, more simple comparison
\cite{nptom}.
In this model of a primary r--process, a blob of matter at high
temperature ($T_9=9$) expands and cools.
Due to the initially high $T$, all
reactions, including charged-particle reactions, are in equilibrium and
the resulting abundances are determined by NSE.
The charged-particle reactions, in
particular the $\alpha$ captures, cease at time $t_\alpha \approx 0.08$ s
at around $T_9 \approx 2.5$.
Below that temperature
a simpler network can be employed, only including (n,$\gamma$),
($\gamma$,n), and $\beta$--decays (subsuming $\beta$--delayed neutron
emission). The seed abundances for this
r--process network are given by the freeze-out abundances of the 
charged-particle network. If the 
triple--$\alpha$ rate is too slow to convert
all $\alpha$'s to heavy mass nuclei at the given charged-particle
freeze-out conditions, an $\alpha$-rich
freeze-out is found. 
The process conditions are specified by the entropy $S$, the
electron abundance $Y_\mathrm{e}$, and the expansion timescale $\tau$.
The number of free neutrons available for capture after the charged
freeze-out also depends on these conditions.
Due to the high $T$, we still find an
(n,$\gamma$)--($\gamma$,n) equilibrium at $t_\alpha$. The
$\beta$-halflives of the most abundant nuclei in each isotopic chain
(these are only one or two due to the shape of the equilibrium equation)
determine how fast material can be converted to the next element. Each
chain remains in equilibrium until time $t_\mathrm{e}$, when the neutron
reactions become too slow to maintain it. Later, at time
$t_\mathrm{fo}$, finally the neutron reactions fully freeze out. Thus,
neutron capture rates are only relevant for $t_\mathrm{e}\leq t\leq
t_\mathrm{fo}$.

Since the uncertainties in the neutron capture rates might be large,
a few exemplary cases are shown here: with standard rates
and with neutron captures multiplied by a factor of 100 and a factor of 0.01,
respectively (this implies that the photodisintegrations are changed by
the same factor). 
The same expansion was chosen as used by \cite{frei99} in their case of 
$\tau=50$ ms.
Assuming $Y_\mathrm{e}=0.45$, for $S\leq 140$ no or only few free neutrons
are left after $t_\alpha$ \cite{frei99}. With such
entropies peaks in the mass range $A\leq 110$ are directly produced from
NSE abundances (slightly modified by final $\alpha$ captures), favoring
nuclei ($Z$,$A$) with $Z$/$A$=$Y_\mathrm{e}$ and high binding energies, but
neutrons do not play a role yet. They are only involved in reaching 
the third r--process peak at higher entropies.
For two entropies $S=240, 310$ Fig.\ \ref{fig:nnabun} presents 
the $n_\mathrm{n}$ as a function of time and the
final abundances. It was previously shown
\cite{frei99} that the freeze-out at higher entropy is slower and that
final neutron captures can alter the resulting abundances of heavy
nuclei but not of light ones. The trough before the high-mass peak
was filled by late neutron captures.

The freeze-out behavior obtained here depends
on the chosen neutron rates. The time at which the $n_\mathrm{n}$ for
the three cases diverge in Fig.\ \ref{fig:nnabun} indicates $t_\mathrm{e}$.
After this point it depends on the entropy how far up in
mass nuclei have been produced and on the final neutron captures how their
abundances are altered. As can be seen in the figure,
$t_\mathrm{fo}$ occurs earlier for larger rates, with $0.2\leq
t_\mathrm{fo}\leq30$ s.
This reflects the increased capture at $t\geq t_\mathrm{e}$.
Contrary to the other cases, r--processing with the slowest rates ceases
due to low temperatures, not due to lack of neutrons.
Nuclei with $A$$\geq$$140$
are mainly produced at $t\geq t_\mathrm{fo}$ and are
therefore more sensitive to the value of the neutron captures.
Especially in the high entropy case it is
evident that faster neutron captures smooth the abundance distribution
and fill the trough before the $A\approx 200$ peak. Due to the longer
duration of the neutron captures, the third peak is shifted to higher
$A$ for the slow case.

\begin{figure}[t]
\begin{center}
\mbox{
\subfigure{\includegraphics[width=4cm,angle=-90]{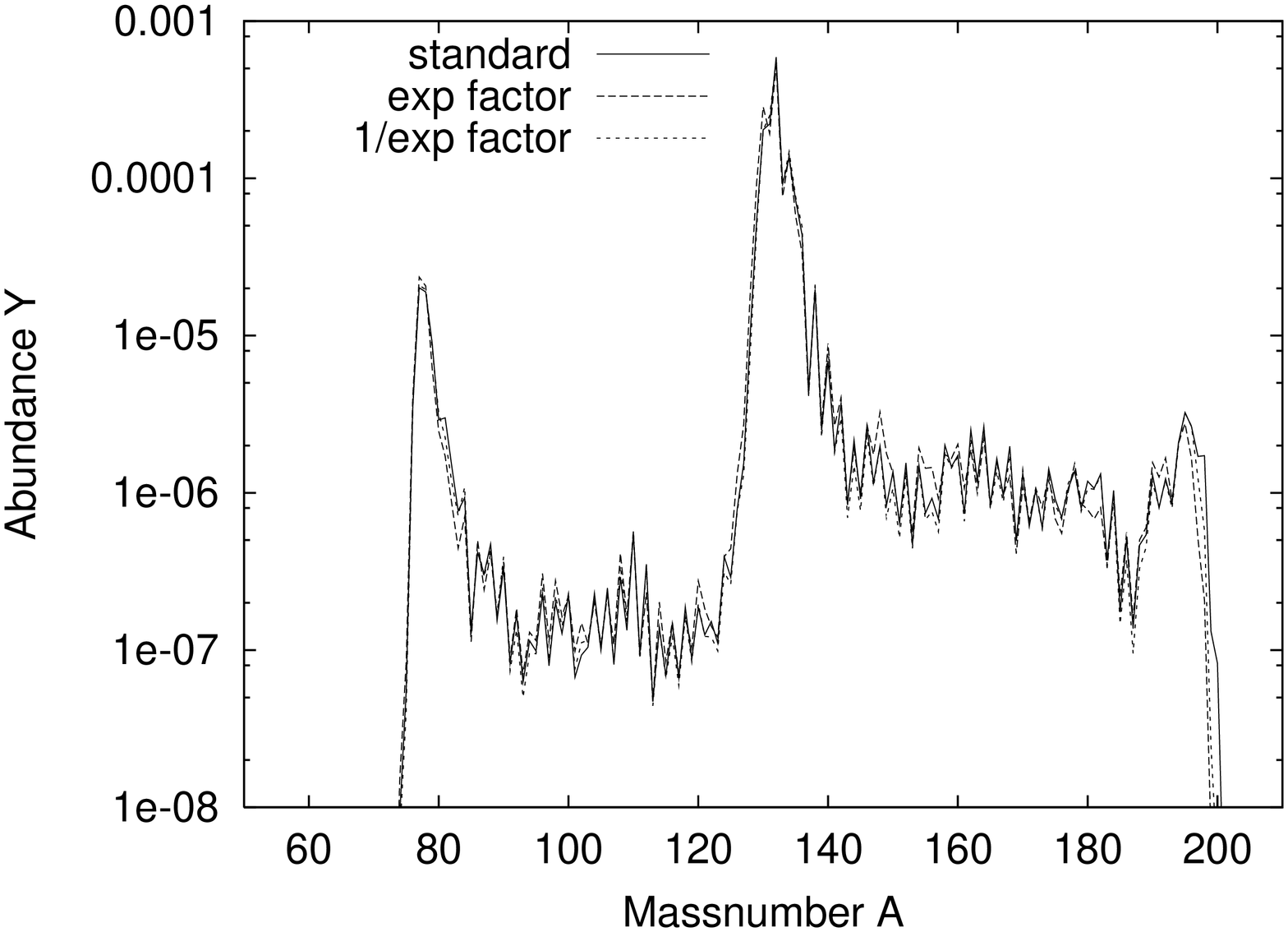}
}
\subfigure{\includegraphics[width=4cm,angle=-90]{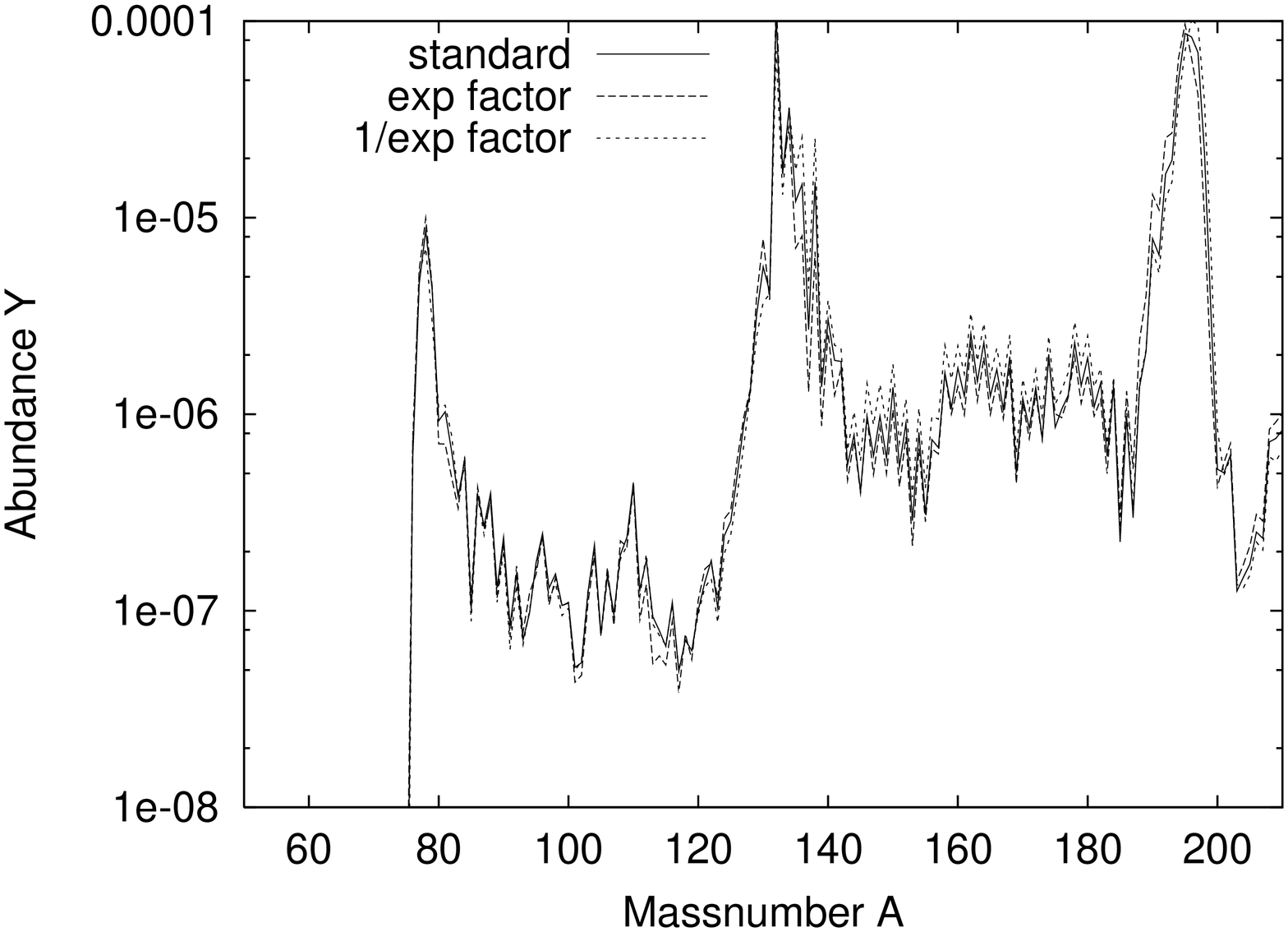}
}
}
\caption{\footnotesize Abundances resulting from an $S_\mathrm{n}$--dependent 
rate variation for 
$S$=240 (left), 310 (right). Shown
are standard rates (full lines), rates enhanced (dashed),
and rates suppressed (dotted) at low $S_\mathrm{n}$.}
\label{fig:exp}
\end{center}
\end{figure}

Despite the fact that there might be
considerable uncertainties in the theoretical rates far off stability
changing {\it all} rates in a range of 4 orders of magnitude seems
unrealistic. Even if new effects (like pygmy resonances \cite{gor,sgor} or
overestimated cross sections \cite{desrau04,sgor})
might change the rates by factors of $10-100$ for extremely neutron-rich
nuclei, late-time captures will not include such nuclei but will occur
closer to stability. Therefore we also used a more realistic variation
of the rates as an exponential inversely depending on 
$S_\mathrm{n}$,
thereby simulating the possible enhancement by pygmy resonances.
Also accounting for the fact that the statistical model
cannot be applied for low $S_\mathrm{n}$, the resulting overestimate was
simulated by using the same function but dividing the rates instead of
multiplying them with the correction factor. This factor reaches a value
of about 100 towards $S_\mathrm{n}\approx 0$ but falls off quickly and is
essentially unity already at $S_\mathrm{n}\approx 3$ MeV \cite{gor}. Fig.\
\ref{fig:exp} shows that there is little impact on the resulting
abundances since only few neutron reactions will occur at low
$S_\mathrm{n}$.

Concluding, 
only large modifications of neutron rates lead to appreciable changes in
the final abundances due to the short relevant time-scale $\Delta
t=t_\mathrm{fo}-t_\mathrm{e}$.
The simple comparison shown above for the hot bubble model has to be
interpreted cautiously.
For reproducing the solar r--process
pattern it is necessary to superpose a number of components with
different entropies. Thus, effects of rates altered on a large scale, as
shown above, can be compensated by a scaling in entropy and a different
weight distribution. Details will be discussed in a forthcoming,
extended paper.
Despite the above caveats the main conclusions are consistent with other
studies \cite{frei99,sur}.
Components with high entropy freeze out slower and late-time
neutron captures can modify the final abundance distribution mainly in
the region $A>140$. Therefore, emphasis has to be put on improving the
prediction of nuclear cross sections and astrophysical reaction rates in
that mass region far from stability.

\end{document}